\documentclass[prl,twocolumn]{revtex4}
\usepackage{graphicx}
\usepackage{dcolumn}
\usepackage{amsmath}
\usepackage{bm}
\newcommand{\be}{\begin{equation}}
\newcommand{\ee}{\end{equation}}
\newcommand{\ba}{\begin{eqnarray}}
\newcommand{\ea}{\end{eqnarray}}
\begin{document}
\title{\emph{Ab initio} molecular dynamics study of\\dissociation of water under an electric field}
\author{A. Marco Saitta$^1$~\cite{aff1}, Franz Saija$^2$~\cite{aff2}, Paolo V. Giaquinta$^3$~\cite{aff3}}
\affiliation{
$^1$ IMPMC, CNRS-UMR 7590, Universit\'e P \& M Curie, 75252 Paris, France \\
$^2$ CNR-IPCF, Viale Ferdinando Stagno d'Alcontres 37, 98158 Messina,
Italy\\
$^3$ Universit\`a degli Studi di Messina, Dipartimento di Fisica, Contrada
Papardo, 98166 Messina, Italy }
\date{\today}
\begin{abstract}
The behavior of liquid water under an electric field is a crucial phenomenon in science and engineering. However, its detailed description at a microscopic level is difficult to achieve experimentally. Here we report on the first \emph{ab
initio} molecular-dynamics study on water under an electric field. We observe that the hydrogen-bond length and the molecular orientation are significantly modified at low-to-moderate field intensities. Fields beyond a threshold of about $0.35$~V/{\AA}~are able to dissociate molecules and sustain an ionic current via a series of correlated proton jumps. Upon applying even more intense fields ($\sim 1.0$~V/{\AA}), a $15-20\%$ fraction of molecules are instantaneously dissociated and the resulting ionic flow yields a conductance of about $7.8~\Omega^{-1}\cdot cm^{-1}$, in good agreement with experimental values. This result paves the way to
quantum-accurate microscopic studies of the effect of electric fields on
aqueous solutions and, thus, to massive applications of \emph{ab initio} molecular
dynamics in neurobiology, electrochemistry and hydrogen economy.
\end{abstract}
%\pacs{}
%\keywords{}
\maketitle

Water is the most abundant compound on Earth and plays a fundamental role
in the realm of natural and life sciences, at the interface between a
variety of disciplines such as physics, chemistry and
biology~\cite{ben,ball}, with a seemingly infinite range of technological
applications. Despite the interest and the enormous amount of fundamental
and applied research work, a complete understanding of its unique behavior
is still lacking~\cite{ball2}. The peculiar properties of water are due to its
molecular structure and dipole moment and, as a consequence, to its
hydrogen-bonding properties~\cite{gerstein}. Among the physical-chemical phenomena of greater interest, particularly relevant is the phenomenon of autoprotolysis, \emph{i.e.} the dissociation of a water molecule according to the following reaction~\cite{iupac}:

\be {\rm{2{H_{2}O}}} \rightleftharpoons {\rm{OH^{-} + H_{3}O^{+}}} \label{eq1} \ee

\noindent where two molecules form a pair of
hydronium and hydroxide ions which then separate, thus determining the pH of
the substance. This effect is of paramount importance because it is a key step in
all processes in which water is involved as more than just a passive
solvent. The kinetics of the autoionization process has
been studied in some experiments~\cite{natzle}, but it is an extremely rare
reaction which takes place on the femtosecond time scale of molecular motions~\cite{eigen},
thus making its microscopic investigation very difficult. On the other
hand, a deeper understanding of this reaction is crucially important in the presence of an external electric field, with consequences in disparate domains, from neurobiology~\cite{kaila} to electrolytic batteries~\cite{reddy} and hydrogen-based technology~\cite{zoulias}. For
example, the activity of neurons is governed by the exchange of ions through
the cell membrane, which is driven by the potential difference between the interior of
the cell and its outside environment. The first experiments of forward and backward
conversion of chemical energy into electric energy date back to 1789; the
first battery was invented by Volta in 1800 and then
used to carry out, for the first time, the electrolysis of water. Since then,
electrochemistry has become a stand-alone branch of science and technology,
with a huge impact on everyday life and more to come on the nascent
hydrogen economy.

The chemical processes governing the autoprotolysis of water in the
absence of an external field are in principle within the reach of
\textit{ab initio} molecular-dynamics simulations. However, because of the
large timescales involved, a straightforward molecular-dynamics approach
turns out to be rather ineffective. The problem was overcome using a
combination of Car-Parrinello (CP) molecular dynamics~\cite{cp} and transition-path sampling~\cite{bolhuis}, which confirmed that water dissociation is an
extremely unfavored event from the energetic point of view. In a
milestone study Geissler \emph{et al.} showed that the rare autoprotolysis of a water molecule is induced by occasional large fluctuations of the electric field generated by the
surrounding solvent, \emph{i.e.} by other water molecules~\cite{geissler}. To test their
theory, these authors adopted a clever scheme in which the occurrence of such rare
fluctuations was favored by \emph{ad hoc} semi-random moves of protons
along hydrogen bonds; they showed that this method is equivalent to applying an external electric field and proved that such a field can dramatically change the energetic balance of the dissociation
reaction, favoring the occurrence of the latter. The nascent hydronium and
hydroxide ions then separate along a chain of hydrogen bonds via Grotthuss
transfer events~\cite{grotthuss}. 
It also turns out that their recombination is made possible through
a collective compression of the water wire bridging the
ions~\cite{hassanali}.
However, when the hydrogen-bond chain still connecting the two ions breaks, 
their rapid recombination is inhibited and from
such a state ions can move further apart and separate completely.
Based on this result, recent classical molecular-dynamics simulation studies have
shown that the important contributions to the electric field acting on OH
bonds stem from water molecules less than 7 {\AA}~away, corresponding to
charges on water molecules of the first and second hydration
shell~\cite{reischl}, and that they are able to generate instantaneous
fields up to $2.5-3.0$ V/{\AA}.

Despite the major progress that such works represent in this
domain, it has not been possible to establish, so far, a quantitative relation
between the degree of water
dissociation and the intensity of the electric field. Several studies based
on the classical molecular dynamics of water molecules under an external
electric field are already present in the literature, describing, for example, the
effect of an electric field on the phase diagram~\cite{aragones} or the phenomenon
of electrofreezing~\cite{kusalik,wei}. However, such studies are typically
based on rigid-molecule non-polarizable potential models, which are unable to
describe the dissociation of water molecules. On the other hand, the
effect of an electric field in \emph{ab initio} Density Functional Theory
(DFT) codes is traditionally taken into account through a sawtooth
potential that is quite satisfactory in low-dimensional systems but which cannot
be used in bulk systems. We present here the first, to our knowledge, \emph{ab initio} study of the properties of bulk water in the
presence of intense external electric fields. The study was carried out using \emph{Quantum Espresso}~\cite{giannozzi},
a well-known open-source package for the DFT calculation of electronic-structure properties,
which uses a plane-wave basis set and pseudopotentials. Macroscopic polarization and finite electric fields are
treated by this code with the modern theory of polarization (Berry phases)~\cite{umari}.
Previous \emph{ab initio} calculations of water in an external field were typically carried out on small
clusters in the vacuum~\cite{choi}. All the technical details of the calculation are reported in the Supplementary
Material file~\cite{SuppFile}.

\begin{figure}
\includegraphics[width=10cm]{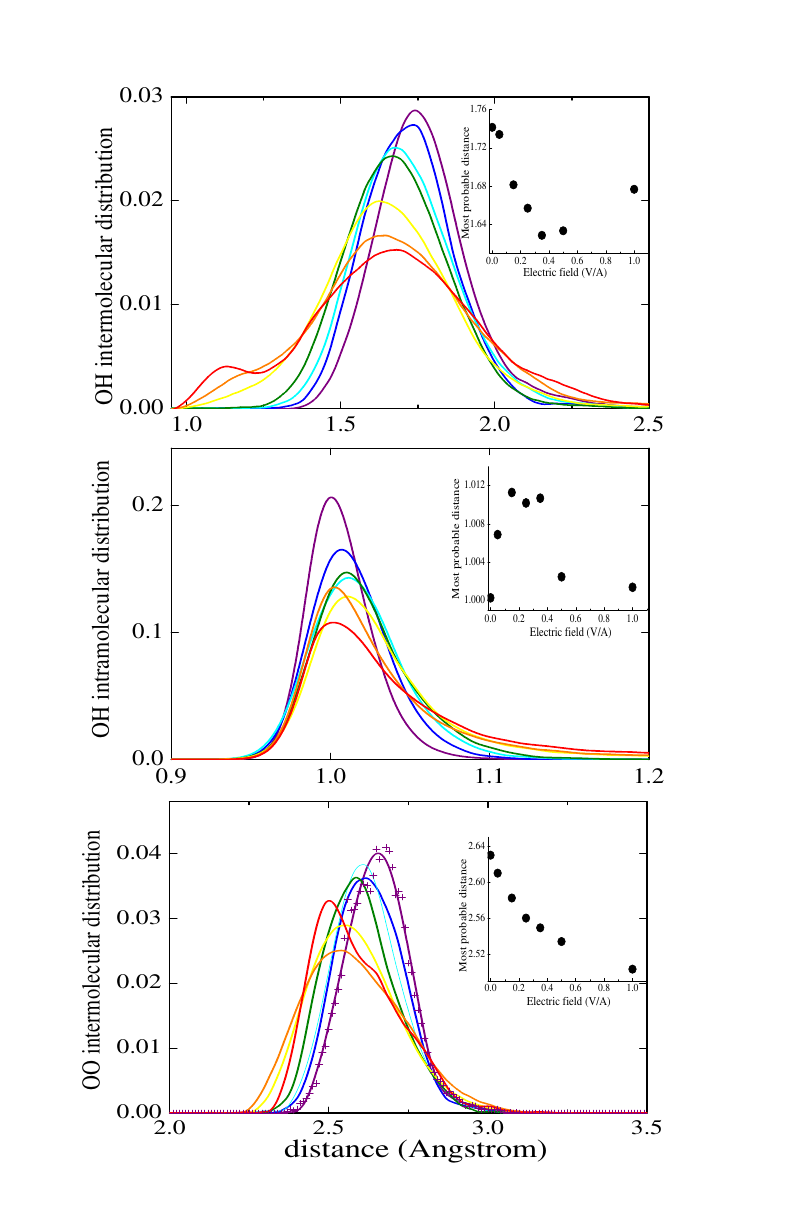}
\caption{Distributions of istantaneous $\rm{O-O}$ and $\rm{O-H}$ distances for electric fields in the range between zero (purple curves) and 1 $\rm{V/{\AA}}$ (red curves). The insets show the evolution of most probable distances with the field intensity.  Top panel: intermolecular distance between an oxygen atom and its third hydrogen neighbor; central panel: intramolecular distance between an oxygen atom and its second hydrogen neighbor; bottom panel: oxygen-oxygen closest-neighbor distance. The experimental data were smoothed with the Savitzky-Golay filter method, as typically shown for the zero-field curve in the bottom panel.}
\label{fig1}
\end{figure}

In order to test the reliability of the method, we first checked the effect of a uniform electric field by calculating the field-induced electronic polarization at fixed atomic positions
for several water configurations as in Ref.~\cite{umari}; the estimate we obtained
$(\epsilon_{\infty}\approx 1.74)$ is in good agreement with the experimental data as well as with previous theoretical results  (we refer the reader to Ref.~\cite{SuppFile} for more detailed information). We also calculated the structural properties of the system at zero field, such as the radial distribution functions,
and found that they reproduce with fair accuracy those reported in the literature~\cite{SuppFile}.
As shown in Ref.~\cite{saitta}, the evolution of the structure of water and of its H-bonding properties are better described
by the instantaneous distribution of the distance between an oxygen atom and its third hydrogen neighbor,
\emph{i.e.} typically the closest H neighbor covalently bonded to a nearby oxygen atom.
We report such a distribution, for values of the external electric field up to 1~V/{\AA}, in the top panel of Fig.~\ref{fig1},
together with the intramolecular $\rm{O-H}$ distance distribution of the second hydrogen neighbor (central panel),
and the $\rm{O-O}$ closest-neighbor distance distribution (bottom panel).
The effects produced by increasing electric fields on the first distribution (top panel) are a systematic broadening
and a significant shortening of the H-bond for fields up to about $0.35$~V/{\AA}; for larger intensities,
the maximum of the distribution weakly shifts again to larger and larger values while a secondary peak emerges at
about 1.1 {\AA}, which can be associated to a covalently bonded hydrogen in a $\rm{H_{3}O^{+}}$ ionic species.
The effect of the field is also manifest on the intramolecular covalent bond between an oxygen atom and its second
hydrogen neighbor, as shown in the central panel of Fig.~\ref{fig1}: the associated distribution becomes more and more
delocalized while the most probable distance exhibits a nonmonotonic trend, opposite to that observed above for the
hydrogen bond. Note also, in the bottom panel of Fig.~\ref{fig1}, the monotonic shortening of the $\rm{O-O}$ distance
up to an amount of about 0.14 {\AA}, a shift comparable to that observed in water under a pressure of about
20 GPa~\cite{schwegler}.
In classical simulations intermolecular distances are unaffected while an increased number of H-bonds has been observed with
longer lifetimes~\cite{wei}. We interpret this partial discrepancy as a clear indication of a larger,
field-induced $\rm{O-H}$ intramolecular polarization, which strengthens the intermolecular $\rm{O-H}$ interaction
and which cannot be accounted for in the classical models most commonly used to describe water. We also observed, indeed rather unsurprisingly as this effect has been already predicted by classical simulations~\cite{wei}, that water
molecules tend to align their dipoles along a direction parallel to the field~\cite{SuppFile}.

\begin{figure}
\includegraphics[width=8.5cm]{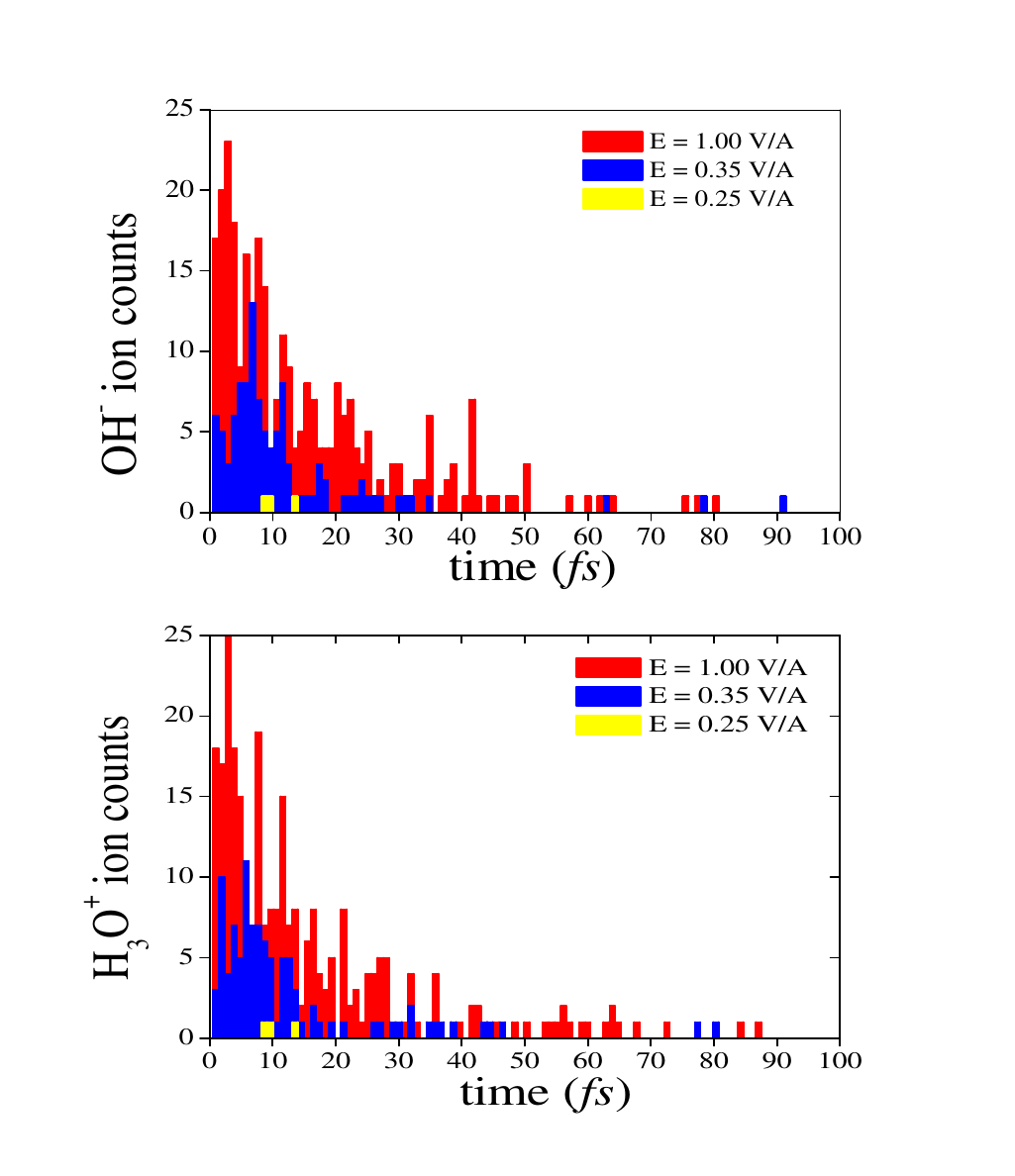}
\caption{Counts of $\rm{OH^{-}}$ and $\rm{H_{3}O^{+}}$ ionic species with lifetimes as absciss\ae, for field intensities 0.25 V/{\AA} (yellow), 0.35 V/{\AA} (blue), and 1 V/{\AA} (red).}
\label{fig2}
\end{figure}

According to the present calculations, water molecules start to dissociate for field intensities around 0.25 V/{\AA}. Protons participating in H-bonds, mostly aligned with the external field, start to jump occasionally back and forth along the bond, thus creating instantaneous $\rm{H_3O^+-OH^-}$ pairs. Such events are still rare and produce extremely short-lived ($\lesssim 10 fs$) ionic pairs,
the field being not strong enough to permanently sustain instantaneous ionic dipoles for times long enough to start a real proton diffusion (see Fig.~\ref{fig2}). Only for fields of about $0.35$ V/{\AA} we started to observe much more frequent molecular dissociations,
suggesting that the balance of the reaction described in Eq.~\ref{eq1} more and more favors the products in the right term
at the expense of the reactants in the left term, {\emph{i.e.}} the neutral state.
We note that the above field threshold is in very good agreement with experiments.
In a work of a few years ago~\cite{rothfuss}, water dissociation was observed to occur under external fields of 0.32 and
0.44 V/{\AA}, the threshold depending on the temperature. A very recent experimental study provided a thorough review of the
process of water ionization in high electric fields induced by Pt electrodes~\cite{stuve}.
In particular, an interesting analysis of the interplay between the local field, as estimated by
Geissler \emph{et al.}~\cite{geissler}, and the external field concludes that the latter should enhance water
ionization for intensities of $0.3-0.6$ V/{\AA}, in excellent agreement with our results.

\begin{figure}
\includegraphics[width=8.5cm]{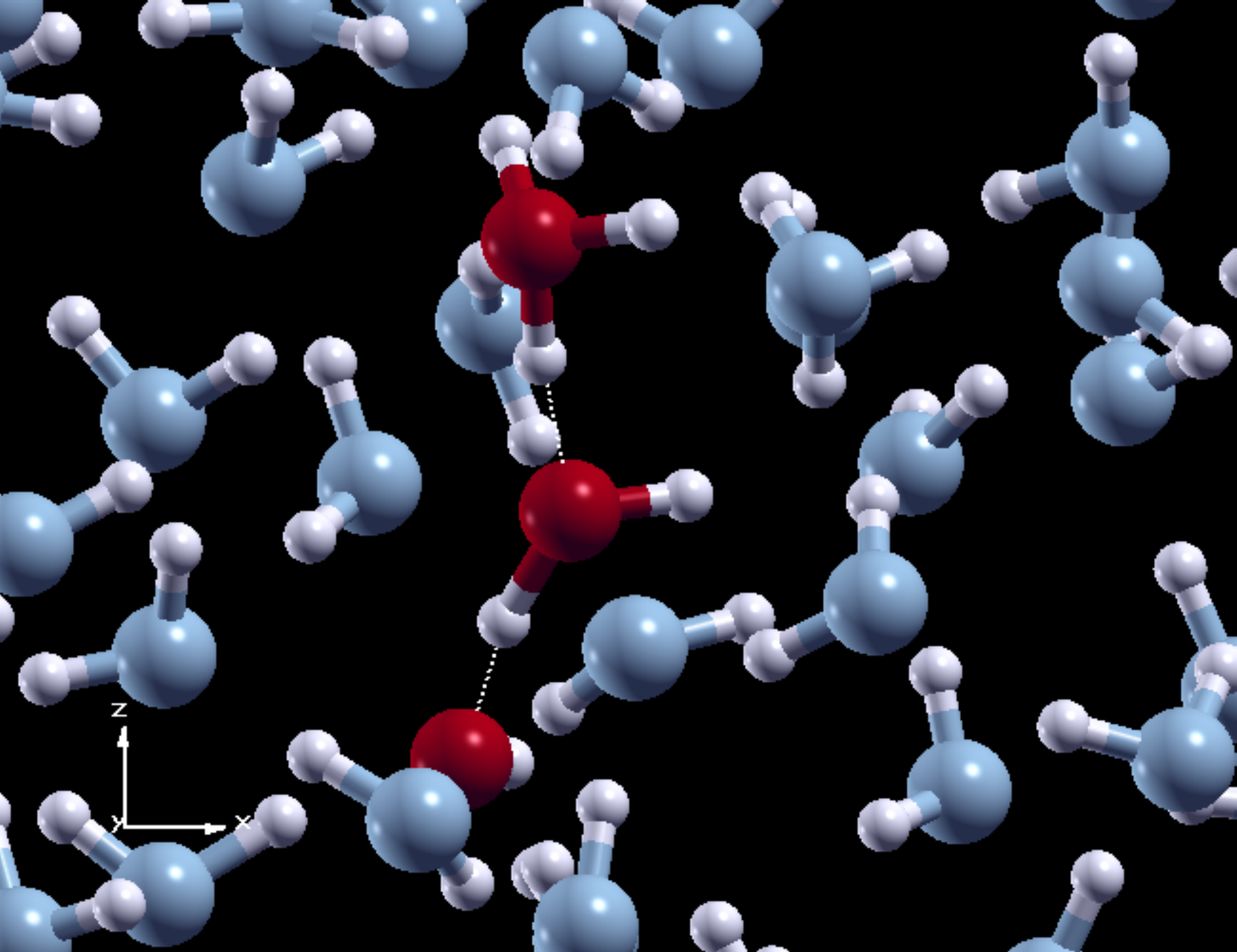}
\caption{Snapshot of a typical dissociation-diffusion mechanism; the molecules
involved in the process are rendered with red coloring, while the surrounding
molecules have gray-colored oxygen atoms. In both cases, hydrogens are
depicted in white. In particular, a $\rm {H_{3}O^{+}-H_{2}O-OH^{-}}$
entity is clearly visible and its relevant hydrogen-bonds are indicated with a dashed
white line. The electric field is oriented along the positive $z$ direction.}
\label{fig3}
\end{figure}

\begin{figure}
\includegraphics[width=9cm]{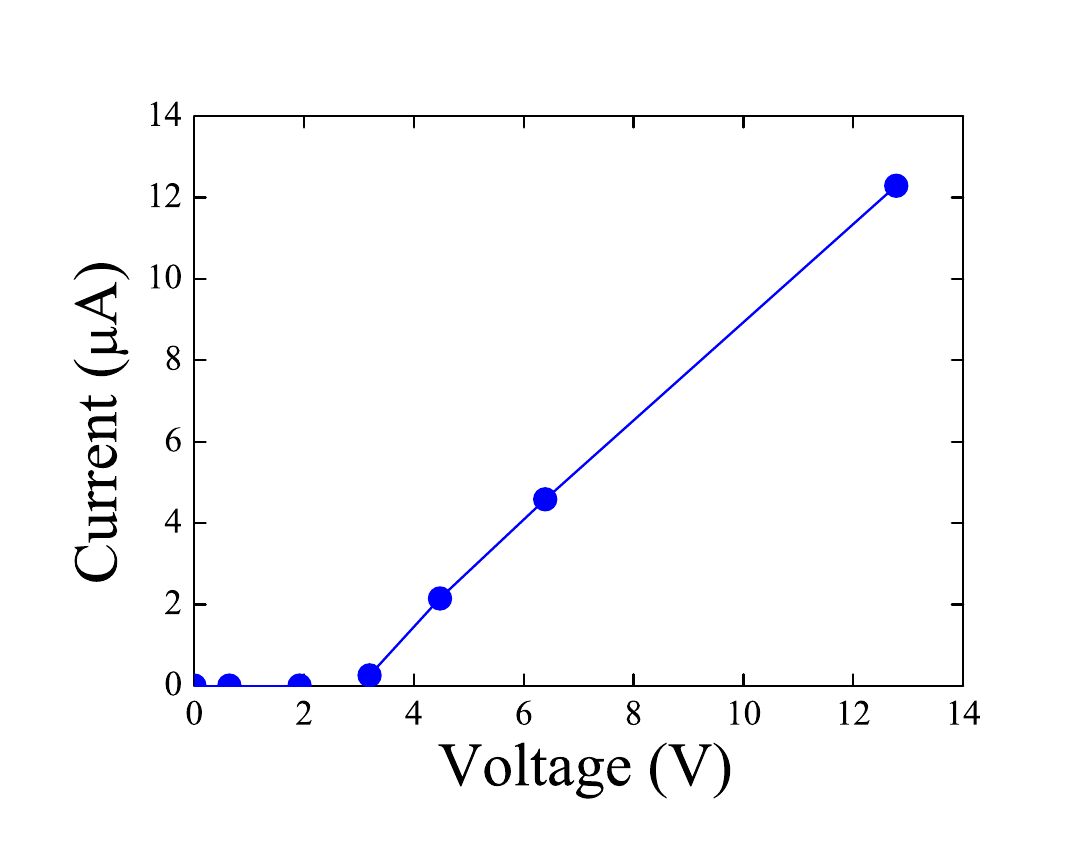}
\caption{Ionic current-voltage characteristic for a cubic cell of $12.42$ {\AA}~side.
The dots represent the calculated values, while the solid line is a guide for the eye.}
\label{fig4}
\end{figure}

Once the external field is strong enough to induce a significant dissociation, ionic lifetimes become dramatically longer, ranging from a few femtoseconds to more than 0.1 $ps$ for both $\rm{H_{3}O^{+}}$ and $\rm{OH^{-}}$, as reported in Fig.~\ref{fig2}. Indeed, at this point the dissociation reactions are typically followed
by a series of proton jumps from the newly formed hydronium to the adjacent H-bonded neutral molecule along the field direction,
forming a new hydronium from which a (different) proton can jump, and so on and so forth.
In other words, ionic diffusion (and thus conduction) is activated by the dissociation reaction described in Eq.~(\ref{eq1}),
but the presence of the field breaks the symmetry of the physical space, providing a spatial direction for the relevant proton hopping events, and thus favoring (or disfavoring) their respective reaction equilibria according
to the relative physical position of the reactants, with respect to each other and to the field.
This effect can be visually pictured as follows:
\be {\rm{H_{2}O + H_2O}} \xrightarrow[E]{~~~} {\rm{OH^{-} + H_{3}O^{+}}}, \label{eq1b} \ee
where the right-directed oriented field induces a proton jump from the left water molecule
to the right one. By analogy, beyond the dissociation threshold the recombination processes
more likely occurs as:
\be {\rm{H_{3}O^{+} + OH^{-}}} \xrightarrow[E]{~~~} {\rm{H_{2}O + H_2O}}, \label{eq1c} \ee
{\emph{i.e.}} the extra proton comes from an hydronium in the opposite direction of the hydroxide
with respect to the field.
Besides these two steps, ionic conduction is sustained by other processes which can be decomposed in terms of 
the following elementary proton jumps:
\be {\rm{H_{3}O^{+} + H_{2}O}} \xrightarrow[E]{~~~} {\rm{H_{2}O + H_{3}O^{+}}} \label{eq2} \ee
and
\be {\rm{H_{2}O + OH^-}} \xrightarrow[E]{~~~} {\rm{OH^{-} + H_{2}O }} \label{eq3}, \ee
where, again, proton hopping along the field direction is privileged.
We note that proton jumps are significantly correlated; in particular, protons 
originally forming donor H-bonds and belonging to adjacent molecules jump almost simultaneously
as their respective acceptor molecule, in a Zundel-to-Zundel mechanism very similar to the one described
in {\it ab-initio} simulations~\cite{geissler} as well as in experiments~\cite{roberts}.
The mechanism of correlated proton jump diffusion is illustrated by a snapshot of our simulation cell in Fig.~\ref{fig3}.
Dissociated molecules recombine with protons diffusing from neighboring molecules, thus contributing to the global
liquid conductivity. We observed that even for fields of $1.0$ V/{\AA} adjacent $\rm{H_3O^+-OH^-}$ pairs are not viable,
unless they are separated by an $\rm{H_2O}$ molecule, and that the Zundel-to-Zundel diffusion mechanism is dominant.
Such correlated processes can involve even more than two protons at the same time, but still
the decomposition of the diffusion mechanism in the individual elementary steps of Eqs.(\ref{eq1b}-\ref{eq3}) is possible,
and allows a more quantitative description of the whole phenomenon and of the corresponding rates. 
To this end, we reported in Table~\ref{tab1} the number of such elementary events per molecule and per picosecond.
We note that at the dissociation threshold (0.35 V/\AA) a single molecule
undergoes, on average, about 3 proton-hopping events in a timespan of 1 picosecond,
about 6 at 0.5 V/\AA, up to about 10 at the highest field considered in this work,
and that the fraction of instantaneously ionized molecules $\Delta N$ increases accordingly.
\begin{table}[htbp]
\begin{center}
%\begin{tabular*}{8cm}{@{\extracolsep{\fill}}clcccc}
\begin{tabular}{cccccc}
\hline
\hline

Field (V/\AA)~    & ~Eq.(\ref{eq1b})~  & ~Eq.(\ref{eq1c})~ & ~Eq.(\ref{eq2})~ & ~Eq.(\ref{eq3})~  & ~$\Delta N$~ \\

\hline

0.25                       &     0.148        &    0.135        &     0.067      &     0.108        &     0.0046       \\
0.35                       &     0.952        &    0.872        &     0.743      &     0.823        &     0.0408       \\
0.50                       &     1.211        &    1.049        &     1.808      &     1.856        &     0.0828       \\
%0.75                       &                  &                 &                &                  &                  \\
1.00                       &     2.341        &    2.082        &     2.502      &     2.728        &     0.1556       \\

\hline
\hline
\end{tabular}
\caption{Calculated event rates and ion fractions for different field intensities,
per molecule and per picosecond of dynamical evolution:
dissociation (Eq.(\ref{eq1b})); recombination (Eq.(\ref{eq1c})); hydronium-to-water hopping (Eq.(\ref{eq2}));
water-to-hydroxide hopping (Eq.(\ref{eq3})). The average fraction of ionized molecules is reported in the last column.
}
\label{tab1}
\end{center}
\end{table}
In particular, at the most intense field a fraction $\Delta N\approx 0.15-0.20$ of the total number of molecules
are instantaneously ionic, as appears from the rightmost column of Table~\ref{tab1}, giving rise to a ohmic
conductivity of about $7.8~\Omega^{-1}\cdot cm^{-1}$; this latter finding is in fairly good agreement with the
available experimental values~\cite{hamann,chau1,chau2} of water dissociation under pressures larger than 20 GPa,
as analogously observed in \emph{ab initio} CP calculations carried out in similar extreme
conditions~\cite{schwegler,cavazzoni}. In particular, the authors of Refs.~\cite{chau1} and \cite{chau2} observed in
shock-wave experiments a conductivity of about $30~\Omega^{-1}\cdot cm^{-1}$, which they attributed to a
massive presence of ionic species; they also estimated that fully ionized water (\emph{i.e.} 100 \% of
dissociated molecules) should have, at those regimes, a conductivity of about $36~\Omega^{-1}\cdot cm^{-1}$.
Our result thus seems to indicate that strong electric fields, yet similar to those instantaneously generated in
neutral water by the surrounding molecules~\cite{reischl}, are able to generate ionic species and to induce an
ionic current whose intensity is comparable to that observed in different experimental conditions and setups.
The calculated ionic current-voltage characteristic of water is finally presented in Fig.~\ref{fig4}.

We conclude that the \emph{ab initio} Car-Parrinello approach is able to catch the essential features of molecular
dissociation and diffusion mechanisms of bulk water under the action of an intense electric field.
The results illustrated in this Letter show that the method employed not only efficaciously describes the
dissociation of the molecule but further accounts for the presence of the forming ionic species whose effect,
even if short lived, is very important in the dynamics of ions in solution. It appears that \emph{ab initio}
simulations of aqueous/molecular systems under an electric field are now feasible with a numerical accuracy
which may already allow for a comparison with the available experimental data. A stimulating scenario of new
investigations opens up with a range of applications of potentially enormous reach, from planetary science, to
biology, electrochemistry and the fast-growing field of hydrogen-based economy.

\section*{Acknowledgements}
We acknowledge the IDRIS Supercomputing Facility for CPU time (CP9-101387).

\end{document}